%% file: main.tex
\PassOptionsToPackage{numbers,sort&compress}{natbib}
\documentclass[preprint,12pt, a4paper]{elsarticle}

\usepackage[dvipsnames,table,xcdraw]{xcolor}
\newcommand{\reviewerOne}[1]{{#1}} % orange
\newcommand{\reviewerTwo}[1]{{#1}} % blue
\newcommand{\reviewerThree}[1]{{#1}} % \textcolor{black!20!green}

\newcommand{\reviewerFour}[1]{{#1}}

\title{\reviewerOne{Flow Gym: A framework for the development, benchmarking, training, and~deployment of flow-field quantification methods}}
\author[eth]{Francesco Banelli\fnref{eqcontrib}}
\author[eth]{Antonio Terpin\fnref{eqcontrib}}
\author[eth]{Alan Bonomi}
\author[eth]{Raffaello D'Andrea}
\address[eth]{\reviewerFour{Institute for Dynamic Systems and Control, ETH Z\"urich, Z\"urich, Switzerland}}
\fntext[eqcontrib]{Equal contribution.}
\journal{SoftwareX}
\usepackage{graphicx}

\usepackage{amsmath,amssymb,mathtools,bm,amsthm,bbm,dsfont,stmaryrd,aligned-overset,physics}
 
\theoremstyle{remark}

\usepackage{booktabs}
\usepackage{multirow}
\definecolor{lightgray}{gray}{0.97}
\rowcolors{2}{white}{lightgray}
\usepackage{adjustbox}
\usepackage{tcolorbox}
\usepackage{enumitem}
\usepackage{mdframed}
\usepackage{amsthm}

\usepackage{wrapfig}
\usepackage{subcaption}

\usepackage{tikz,pgfplots}
\usetikzlibrary{patterns,hobby,calc,intersections,spy}
\pgfplotsset{compat=1.18}
\usepgfplotslibrary{statistics,fillbetween}
\usepackage{pgfplotstable}

\usepackage[acronym]{glossaries}
\glsdisablehyper
\newacronym{ccd}{CCD}{Charge Coupled Device}
\newacronym{piv}{PIV}{Particle Image Velocimetry}
\newacronym{epe}{EPE}{end-point error}
\newacronym{rl}{RL}{reinforcement learning}
\newacronym{cv}{CV}{computer vision}
\newacronym{dis}{DIS}{dense inverse search}
\newacronym{cfd}{CFD}{Computational Fluid Dynamics}
\usepackage{pifont}
\renewcommand{\checkmark}{\textcolor{black}{\ding{51}}}
\newcommand{\notcheck}{\textcolor{black}{\ding{55}}}
\newcommand{\maybecheck}{\textcolor{black}{\textemdash}}

\usepackage{microtype}

\usepackage{hyperref}
\usepackage[capitalize,noabbrev]{cleveref}
\usepackage{url}
\usepackage{amssymb}

\usepackage{listings}
\usepackage{tcolorbox}
\tcbuselibrary{listings}

\crefname{lstlisting}{\reviewerFour{Listing}}{\reviewerFour{Listings}}
\Crefname{lstlisting}{\reviewerFour{Listing}}{\reviewerFour{Listings}}

\definecolor{codebg}{RGB}{247,247,247}

\lstdefinestyle{pythoncode}{
  language=Python,
  backgroundcolor=\color{codebg},
  frame=none,
  basicstyle=\ttfamily\small,
  keywordstyle=\bfseries\color{RoyalBlue},
  commentstyle=\itshape\color{ForestGreen},
  stringstyle=\color{BrickRed},
  showstringspaces=false,
  columns=fullflexible,
  keepspaces=true,
  breaklines=true,
  breakatwhitespace=true,
}

\begin{document}
\begin{frontmatter}
\begin{abstract}
\reviewerOne{
Particle image velocimetry (PIV) and related optical-flow methods are widely used to quantify fluid motion, but their development and evaluation are often hindered by fragmented software, inconsistent interfaces, and limited reproducibility. To address these challenges, we present Flow Gym, a framework for developing, benchmarking, training, and deploying flow-field quantification methods, with a primary focus on PIV. Its core contribution is a standardized interface that allows classical and learning-based algorithms to be integrated, compared, and deployed within a common pipeline. The framework includes JAX implementations and wrappers for existing methods, modular pre-processing and post-processing components, and utilities for training and benchmarking. By leveraging JAX, Flow Gym supports hardware-accelerated execution while remaining interoperable with external implementations from libraries such as OpenCV and PyTorch. It can operate on both synthetic and experimental data and supports the same workflow for offline benchmarking and real-time deployment. Flow Gym is designed to improve reproducibility, reduce barriers to method development, and facilitate the translation of flow-field quantification algorithms from research to experimental settings.
}
\end{abstract}
\end{frontmatter}

\clearpage
\section*{Code metadata}
\begin{table}[htb!]
\begin{tabular}{|l|p{6cm}|p{6cm}|}
\hline
\textbf{Nr.} & \textbf{Code metadata description} & \textbf{Metadata} \\
\hline
C1 & Current code version & v0.1.0 \\
\hline
C2 & Permanent link to code/repository used for this code version & \url{https://github.com/antonioterpin/flowgym} \\
\hline
C3  & Permanent link to Reproducible Capsule & \url{https://github.com/antonioterpin/flowgym/blob/main/src/main.py}\\
\hline
C4 & Legal Code License   & MIT License. \\
\hline
C5 & Code versioning system used & git\\
\hline
C6 & Software code languages, tools, and services used & Python\\
\hline
C7 & Compilation requirements, operating environments \& dependencies & Python 3.10+, JAX 0.6.2+, tqdm 4.67.1, h5py 3.13.0+, ruamel.yaml 0.18.10+, robo-goggles 0.1.7+\\
\hline
C8 & If available Link to developer documentation/manual &\url{https://github.com/antonioterpin/flowgym} \\
\hline
C9 & Support email for questions & aterpin@ethz.ch\\
\hline
\end{tabular}
\caption{Code metadata}
\label{codeMetadata} 
\end{table}
\clearpage
\input{figure_1}
\section{Motivation and significance}
\label{sec:intro}
\reviewerOne{Flow-field quantification from images of tracer particles is a central task in experimental fluid mechanics, with \gls*{piv} and optical flow among the most widely used techniques \cite{willert1991digital}. Over the years, a broad range of methods has been developed for this purpose, including both classical approaches \cite{scarano2001iterativeimage,corpetti2006fluid,westerweel2013particle,astarita2007analysis} and learning-based methods \cite{cai2019dense,manickathan2022kinematic,lagemann2021deep,lee2017piv}.
In recent years, advances in computer vision and deep learning have significantly influenced the design of new flow-field quantification methods \cite{zhu2025pivflowdiffusertransferlearningbaseddenoisingdiffusionmodels,huang2022flowformer}.}

\reviewerOne{However, despite the growing methodological diversity, software support for developing, comparing, and deploying such methods remains fragmented. Implementations are often distributed across different libraries and programming frameworks, rely on incompatible interfaces, and differ in pre-processing, post-processing, and evaluation protocols. As a result, fair comparison, reproducible benchmarking, and practical deployment remain unnecessarily difficult.
By contrast, progress in neighboring fields has been accelerated by shared software abstractions, standardized reference implementations, and efficient execution on modern hardware. In computer vision and reinforcement learning, common interfaces and benchmark-oriented software ecosystems have played a central role in improving reproducibility and enabling systematic comparison \cite{brockman2016openaigym,stablebaselines3,opencv_library}. This is particularly important in settings where performance is sensitive to implementation details and experimental choices \cite{hendersonDeepReinforcementLearning2018}. Flow-field quantification pipelines exhibit a similar sensitivity, which makes standardized and reusable software infrastructure especially valuable.}

\reviewerOne{With the software package presented in this paper, we aim to bring to flow-field quantification the same emphasis on reproducibility, comparability, and practical usability that has become standard in adjacent computational fields in the past decade.}

\begin{mdframed}[hidealllines=true,backgroundcolor=blue!5]
\vspace{-0.25cm}
\paragraph{Contributions}
\reviewerOne{We present Flow Gym, a unified framework for developing, training, benchmarking, and deploying flow-field quantification methods; see \cref{fig:cover}. Our main contributions are:}
\begin{itemize}[leftmargin=*, itemsep=-.25em, topsep=0.5em]
    \item \reviewerOne{a standardized \texttt{Estimator} interface for classical and learning-based flow-field quantification methods;}
    \item \reviewerOne{shared training and evaluation workflows that support repeatable benchmarking and facilitate deployment on real experimental setups;}
    \item \reviewerOne{JAX-native implementations and interoperable wrappers for representative methods, providing accelerator-friendly execution together with compatibility with external libraries such as OpenCV \cite{farneback2003two,weinzaepfel2013deepflow,horn1981determining}, PyTorch \cite{lagemann2021deep}, and OpenPIV \cite{openpiv_python_0_23_4}.}
\end{itemize}

\reviewerOne{Flow Gym supports both consecutive (or recurrent) and independent estimation workflows for flow-field quantification. The same interface can also be used for related estimators beyond \gls*{piv}, such as tracer-particle density estimation.}
\end{mdframed}

\section{Software description and illustrative examples}
\label{sec:flowgym}

Flow Gym is organized as a unified software framework for developing, benchmarking, training, and deploying flow-field quantification methods. \reviewerOne{The framework is centered around the \texttt{Estimator} interface, which provides a common foundation for standardized baselines and for integrating new methods within the same workflow.} The same interface can also be extended to related quantification tasks, such as tracer-particle density estimation\footnote{In this paper, for the sake of clarity, we focus on the flow-field quantification instance of the software package because it is the most widely studied in the literature.}.

\subsection{A unified interface}
\paragraph{Estimator}
The \texttt{Estimator} module defines a unified interface for algorithms that map observations (e.g., the \gls*{piv} image pairs) to quantities of interest. 
To align with JAX's programming model, the \texttt{Estimator} interface is stateless and functional. Each call to the \texttt{Estimator} receives as input the latest observation, the state of the \texttt{Estimator}, and the trainable state, and returns an updated \texttt{Estimator} state together with a dictionary of metrics (see \cref{fig:cover}):
\begin{lstlisting}[style=pythoncode]
new_state, metrics = estimator(image, state, trainable_state)
\end{lstlisting}
In particular, \texttt{new\_state} includes the rolled history and \reviewerOne{the updated PRNG key that is carried to the next iteration to ensure random draws in subsequent calls}, and \texttt{metrics} provides task-specific logging information.
The state of the \texttt{Estimator} captures the run-time context propagated across successive calls. For simple algorithms, this context may be limited only to the current image pair. This is the case, for example, for WIDIM-based algorithms and optical flow algorithms. More advanced flow-field quantification methods may also retain a short-term history, such as previous image pairs, previous estimates, or recurrent internal variables. As a result, the same interface supports ``one-shot'' estimators as well as ``recurrent'' estimators that exploit the temporal correlation across estimates. In the current implementation, the state also comprises a batch of PRNG keys for controlled randomness. When image pairs are processed independently (e.g., during benchmarking on a shuffled dataset), the state can simply be re-initialized before each estimation.
The trainable state captures the long-term parameters of the estimator. For learning-based estimators, this typically includes the model weights, optimizer state, and optimizer transformation. \reviewerOne{Classical estimators such as cross-correlation algorithms do not have trainable parameters; in those cases, Flow Gym simply propagates an empty trainable-state container through the same interface so that classical and learning-based methods can be compared and deployed within a common pipeline.} \reviewerOne{Following JAX's functional programming style, the \texttt{Estimator} call is side-effect free: neither the \texttt{trainable\_state} nor the input observation is mutated in place. An explanation of how to add a learning component to an \texttt{Estimator} instance is provided in ``\nameref{par:training}.''}

In addition to the core call, the class also provides standardized hooks for image processing. On the input side, \texttt{Estimator} supports a configurable sequence of \emph{pre-processing} steps, explained in detail in \cref{subsec:pre}. Each step is applied sequentially before estimation, ensuring consistent data pre-processing across algorithms.
For flow-field quantification algorithms, the \texttt{FlowFieldEstimator} subclass additionally provides a standardized \emph{post-processing} stage; see \cref{sec:post}.
This layered design allows pre-processing, estimation, and post-processing to be specified independently.

\paragraph{Deployment of an estimator}

Estimators can be instantiated directly from configuration files. \reviewerOne{Importantly, deployment is agnostic to the provenance of the input data: once instantiated, the same estimator can be applied to synthetic image pairs, recorded experimental datasets, or live acquisitions from a real setup.} The deployment pipeline in \cref{lst:example-make} has already been adopted in several projects, including real-time fluid control \cite{terpin2025ff}.
\begin{lstlisting}[
  style=pythoncode,
  caption={\reviewerFour{Example usage of \texttt{make\_estimator} for deployment.}},
  label={lst:example-make},
  float,
  captionpos=b,
]
# Define the config (or load from YAML)
estimator_config = {
    "estimator": "dis_jax",
    "estimate_type": "flow",
    "config": {"jit": True, ..., },
}

# Create the estimator and associated functions
(
    trainable_state, 
    create_state_fn, 
    compute_estimate_fn,
    estimator
) = make_estimator(
    estimator_config, image_shape
)

# Choose an image pair of interest (synthetic or experimental)
image_0, image_1 = input_images

# Compute an estimate for an image pair
est_state_0 = create_state_fn(image_0, key0)
est_state_1, metrics = compute_estimate_fn(
    image_1, est_state, trainable_state
)

# If this is a stream of images, rather than an image pair, 
# one can use the estimator recurrently given a new image, image_2
est_state_2, metrics = compute_estimate_fn(
    image_2, est_state_1, trainable_state
)
\end{lstlisting}

\paragraph{Implementing a custom estimator}
\label{paragraph:custom_estimator}
Within this interface, implementing a new estimator reduces to defining a single method: \texttt{\_estimate}. This method specifies how a pre-processed observation, together with the current estimator state and trainable state, is mapped to a new state and a set of metrics. The base \texttt{Estimator} class automatically handles auxiliary tasks such as pre-processing, PRNG-key management, and state-history updates. This allows developers to focus on the algorithm-specific logic of their estimator.

\reviewerOne{In particular, contributors adding a new algorithm to Flow Gym can follow one of two approaches:}

\begin{itemize}[leftmargin=*]
    \item \reviewerOne{\emph{Implement the algorithm natively in JAX.} This consists in writing the algorithm logic directly in \texttt{\_estimate} using JAX, thereby maximizing performance and integration, for example through compatibility with \texttt{jit} and \texttt{vmap}. For instance, we implemented DIS \cite{kroeger2016fast} and RAFT32-PIV \cite{lagemann2021deep} natively in JAX.}
    
    \item \reviewerOne{\emph{Wrap an existing external implementation.} This consists in using \texttt{\_estimate} as a wrapper around an external library (e.g., OpenCV or OpenPIV) or around a model implemented in another framework (e.g., PyTorch). Examples include the OpenCV-based DeepFlow estimator \cite{weinzaepfel2013deepflow}, included through the OpenCV API, and RAFT32-PIV \cite{lagemann2021deep}, whose PyTorch code was ported while providing appropriate attribution and respecting the upstream license. This approach enables benchmarking and deployment in Flow Gym even when the original method is not implemented natively in JAX.}
\end{itemize}

\paragraph{Training an estimator}
\label{par:training}
\reviewerOne{When an estimator contains trainable weights or tunable parameters, Flow Gym provides a common training interface built on top of the \texttt{Estimator} abstraction. In particular, a trainable estimator should implement two methods:
\begin{itemize}
    \item \texttt{create\_trainable\_state}, which initializes the \texttt{trainable\_state} object; by default, this is an empty pytree. This method is called automatically within \texttt{make\_estimator}.
    \item \texttt{create\_train\_step}, which returns a function specifying one optimization step. For example, for a neural-network-based estimator, this typically corresponds to one stochastic-gradient-descent step.
\end{itemize}
\cref{lst:example-train} shows a simplified supervised training loop based on the functional interface of \texttt{Estimator} together with a standard data loader. \cref{lst:example-train-consecutive} shows a simplified sequential training loop for a learnable estimator that exploits the history of observations across multiple steps. For this purpose, Flow Gym provides \texttt{FluidEnv}, a lightweight wrapper around SynthPix \cite{terpin2025synthpix} that exposes temporally consecutive observations through a minimal \texttt{reset}/\texttt{step} interface. Flow Gym also provides more structured versions of this boilerplate code in \texttt{train.py} and \texttt{train\_supervised.py}.}

\reviewerOne{Training is not restricted to synthetic data: the same interface can also be used with experimental image pairs, whether processed independently or as part of a temporal sequence; see also \cref{sec:data-sources}. In such cases, when ground-truth flow is unavailable, one can consider a different method as reference or an unsupervised loss defined directly from the images. For example, by warping one image with the estimated flow and measuring its consistency with the other. Sequential training relies on precomputed or externally generated time-resolved flows, which are often available from offline simulations and public datasets such as \cite{jhtdb,graham2016web,li2008public,ohana2024the}.}

\reviewerOne{Not all estimators integrated in Flow Gym are trainable. Classical methods such as OpenPIV are included as standardized reference implementations for benchmarking and deployment under the same interface as learning-based methods, but are not trained with Flow Gym.}

\begin{lstlisting}[
    style=pythoncode,
    caption={\reviewerFour{Example of training loop with \texttt{Estimator} and a data loader.}},
    label={lst:example-train},
    float,
    captionpos=b,
]
# Initialize trainable state and training step
train_step_fn = estimator.create_train_step()

# Iterate through the dataloader
for imgs1, imgs2, ground_truth in data_loader:

    # Create a new state for each batch
    est_state = create_state_fn(imgs1, key)

    # Functionally update the trainable state
    (
        loss, 
        est_state_new, 
        trainable_state_new, 
        metrics
    ) = train_step_fn(
        trainable_state, est_state, imgs2, ground_truth
    )
\end{lstlisting}

\begin{lstlisting}[
    style=pythoncode,
    caption={\reviewerFour{Example of training loop with \texttt{FluidEnv} and \texttt{Estimator}.}},
    label={lst:example-train-consecutive},
    float,
    captionpos=b,
]
# Instantiate the environment
env = FluidEnv.make(env_config)

# Create the training step function
train_step_fn = estimator.create_train_step()

# Iterate through the episodes
for episode in range(num_episodes):

    # Reset the environment for every new episode
    observations, env_state, done = env.reset(env_state)
    est_state = create_state_fn(observations.images, key)

    while not done.any():
        # Estimator forward pass
        est_state, metrics = compute_estimate_fn(
            observations, est_state, trainable_state)

        # Extract the action (the last estimate)
        action = est_state["estimates"][:, -1]
        # Environment step
        observations, env_state_new, reward, done = env.step(
            env_state,
            action
        )

        # Optimization step
        (
            loss, 
            trainable_state_new, 
            est_state_new, 
            metrics
        ) = train_step_fn(
            est_state, 
            trainable_state, 
            reward
        )
\end{lstlisting}

\subsection{Stable baselines}
\label{sec:benchmarking}
We organize the baseline implementations and comparisons into three categories: pre-processing, processing, and post-processing \cite{stamhuis2014pivlab}. \reviewerOne{Although not exhaustive, this collection covers several of the most commonly used components in the \gls*{piv} image-analysis pipeline. By providing common implementations to build upon, Flow Gym facilitates the development, benchmarking, training, and deployment of new algorithms.}

\subsubsection{Pre-processing}
\label{subsec:pre}
We implement in JAX several image pre-processing techniques commonly used in flow-field quantification pipelines \cite{stamhuis2014pivlab}, including:
\begin{itemize}[leftmargin=*]
    \item \emph{Histogram equalization} \cite{pizer1987adaptive,shavit2007intensity,stamhuis2014pivlab}. For every tile, the pixel intensities are adjusted to span the full range, so that low- and high-exposure regions are processed independently to maximize contrast.
    \item \emph{Intensity high-pass} \cite{jahne2005digital,stamhuis2014pivlab}. A high-pass filter removes low-frequency components associated with inhomogeneous lighting while preserving the high-frequency components associated with tracer particles \cite{stamhuis2014pivlab}.
    \item \emph{Intensity capping (clipping)} \cite{shavit2007intensity,stamhuis2014pivlab}. Many \gls*{piv} algorithms are affected by non-uniform particle brightness. To mitigate this effect, the maximum (and optionally minimum) intensity in an image can be clipped.
\end{itemize}
We also implement other standard operations, including Otsu thresholding, Gaussian blurring, and normalization \cite{jahne2005digital,opencv_library}.
We illustrate in \cref{fig:pre-processing} the effects of the different pre-processing techniques.

\paragraph{Pre-processing configuration and customization} 
The pre-processing pipeline in Flow Gym is configured through a list of dictionaries, each specifying a pre-processing function and its parameters. 
To introduce a new pre-processing step, it suffices to define a function with signature
\begin{lstlisting}[style=pythoncode]
images, state, trainable_state = my_step(
    images, state, trainable_state, param1, ...)
\end{lstlisting}
The parameters ``\texttt{param1, ...}'' are automatically extracted from the configuration file when the pipeline is loaded.
This modular design allows flexible composition of transformations such as normalization, filtering, or denoising. Each pre-processing step receives the input image and the estimator’s \texttt{state} and \texttt{trainable\_state}, which also makes it possible to implement dynamic or learning-based pre-processing. For example, one may include adaptive normalization or diffusion-based refinement \cite{shu2023physics}.
\input{figure_4}
\subsubsection{Processing}

\paragraph{JAX implementations}
We re-implement several optical flow and \gls{piv} methods in JAX, including DIS \cite{kroeger2016fast}, OpenPIV \cite{openpiv_python_0_23_4}, and RAFT32-PIV \cite{lagemann2021deep}. \reviewerOne{These implementations provide accelerator-friendly baselines under the same interface used throughout Flow Gym.}

\paragraph{Integration of existing implementations}
Complementarily, Flow Gym provides wrappers around several existing implementations (e.g., OpenCV algorithms \cite{kroeger2016fast,farneback2003two,weinzaepfel2013deepflow,horn1981determining}, OpenPIV \cite{openpiv_python_0_23_4}, and PyTorch implementations \cite{lagemann2021deep}). \reviewerOne{This allows users to compare external implementations and JAX-native estimators under the same conditions, including methods that are not trained within Flow Gym and methods that do not support end-to-end JAX differentiation.}
\begin{table}[b]
    \centering
    \begin{tabular}{l cc}
        Method & Integrated & Implemented in JAX\\ \hline
            DIS \cite{kroeger2016fast} & \checkmark & \checkmark \\
            Farneback \cite{farneback2003two} & \checkmark & \notcheck \\
            DeepFlow \cite{weinzaepfel2013deepflow} & \checkmark & \notcheck  \\
            Horn-Schunck \cite{horn1981determining} & \checkmark & \notcheck \\
            OpenPIV \cite{openpiv_python_0_23_4} & \checkmark & \checkmark \\
            RAFT32-PIV \cite{lagemann2021deep} & \checkmark & \checkmark\\
            PIV-ADMM \cite{bonomi2025admm}& \maybecheck* & \checkmark \\
        \end{tabular}
    \caption{Algorithms currently available in Flow Gym. ``Integrated'' denotes wrapped external implementations; ``Implemented in JAX'' denotes JAX-native implementations. *: JAX native only.}
    \label{tab:methods}
\end{table}

\subsubsection{Post-Processing}
\label{sec:post}
Following \cite{stamhuis2014pivlab}, we implement data validation, data interpolation, and data smoothing, introducing minor variations to these techniques. The parameters of the post-processing steps are specified analogously to the pre-processing steps, and we similarly allow the integration of (possibly learning-based) custom methods; see \cref{subsec:pre}.
\paragraph{Data validation} We implement the following outlier-detection schemes:
\begin{itemize}[leftmargin=*]
    \item \emph{Constant thresholding velocity filter}. All entries with a velocity magnitude outside a constant range $[u_{\min}, u_{\max}]$ are marked as outliers.
    \item \emph{Adaptive thresholding velocity filter (local/global)}. The entries with a velocity magnitude outside $[\bar{u} - n\sigma_u, \bar{u} + n\sigma_u]$ are marked as outliers, where $\bar{u}$ and $\sigma_u$ are computed either globally over the full field or \reviewerTwo{locally within a square neighborhood of size $N\times N$ centered on each interrogation point. In the results of \cref{fig:validation}, we use $N=3$, corresponding to a $3\times 3$ neighborhood.}
    \item \emph{Universal outlier detection based on the median test} \cite{westerweel2005universal}. We implement the algorithm presented in \cite{westerweel2005universal}.
    By making use of comparator networks \cite{knuth1997art} we achieve sub-ms performance on megapixel images.
\end{itemize}

In \cref{fig:validation}, we illustrate the effects of the different outlier detection schemes on an estimated flow from synthetic images.
\input{figure_5}
\paragraph{Data interpolation}

As in \cite{stamhuis2014pivlab}, we implement in JAX standard interpolation techniques to reconstruct the flow field in regions where data have been marked as invalid by the validation step, including:
\begin{itemize}[leftmargin=*]
    \item \emph{Tile-based averaging}, where each missing vector is replaced by the mean of its valid neighbors within a predefined stencil; this approach is computationally efficient and well suited for isolated outliers, but may oversmooth sharp gradients.
    \item \emph{Boundary-value solver}, where the inpainting problem is formulated as solving Laplace's equation with boundary conditions set by the valid neighboring vectors; this approach is computationally more expensive, but provides more globally consistent reconstructions.
\end{itemize}  
Importantly, custom data-interpolation steps, possibly learning-based, such as diffusion-model-based data interpolation \cite{shu2023physics}, can be easily integrated.

\paragraph{Data smoothing}  
To attenuate residual noise and improve local consistency of the estimate, we provide efficient JAX implementations of several smoothing operators, including \emph{average filtering}, \emph{median filtering}, and \emph{normalized least-squares smoothing} \cite{stamhuis2014pivlab}.

\reviewerThree{
\subsection{Data sources}
\label{sec:data-sources}
A key design choice in Flow Gym is that the \texttt{Estimator} interface is agnostic to the provenance of the input data. In particular, estimators and the shared training and benchmarking utilities operate on a standardized batch interface rather than assuming a specific flow solver, renderer, or acquisition pipeline. In the flow-field quantification setting considered in this paper, this interface consists of image pairs and, when available, the associated two-dimensional flow fields (or a two-dimensional slice of a three-dimensional field).
In our current implementation, this standardized batch interface is provided by SynthPix, which can either (i) generate \gls*{piv} image pairs from prescribed flow fields or (ii) load existing datasets; see \cite{terpin2025synthpix}. Consequently, the same estimator can be developed and benchmarked on synthetic data, offline simulation outputs, or recorded experimental datasets without changing its interface. \reviewerOne{Through a custom adapter, SynthPix can also be coupled to an external physics simulator that computes the flow online, enabling closed-loop setups such as active fluid control.}
\reviewerOne{Importantly, deployment of an \texttt{Estimator} is decoupled from SynthPix: once instantiated, the same estimator can be applied directly to experimental image pairs acquired online, including live streams from real \gls*{piv} setups such as water channels, and can therefore be used in real-time physical fluid-control loops \cite{terpin2025ff}.}
As a result, the same estimator interface can be used with a wide range of data sources, including (a) snapshots exported from physics-based solvers, (b) experimental \gls*{piv} datasets with pre-recorded image pairs, (c) analytically specified vector fields, and (d) flows from other scientific domains or non-physical synthetic fields. This also makes it possible to interface Flow Gym with broader simulation collections such as \emph{The Well} dataset \cite{ohana2024the} and the \emph{Johns Hopkins Turbulence Database} \cite{jhtdb}.
}

\section{Impact}
Flow Gym has already been adopted as a shared software infrastructure across multiple research efforts in flow-field quantification and control. Since its initial development, it has enabled the rapid deployment and iteration of real-time flow-field quantification pipelines in experimental settings with hardware in the loop, as demonstrated in~\cite{terpin2025ff}.
The framework also supported the development and evaluation of new estimation methodologies. For instance, Flow Gym was used to implement, test, and benchmark the consensus ADMM-based flow refinement approach introduced in~\cite{bonomi2025admm}.
In addition, Flow Gym was used to study learning-based estimators subject to hard constraints. The experimental validation of hard-constrained neural networks for flow estimation in~\cite{grontas2025pinetoptimizinghardconstrainedneural} relied on Flow Gym to evaluate constrained and unconstrained models within the same pipeline.
Flow Gym currently supports ongoing research at ETH Z\"urich on adaptive PIV tuning and active fluid control.

\section{Conclusions}
Flow Gym is a unified framework for developing, training, benchmarking, and deploying flow-field quantification methods, with a primary focus on \gls*{piv}. Its core contribution is a standardized estimator interface that enables seamless integration, fair comparison, and practical deployment across both classical and learning-based methods.
Implemented in JAX and interoperable with OpenCV and PyTorch, Flow Gym supports accelerator-friendly execution while remaining compatible with existing software stacks.
We believe the following represent meaningful directions for future work:
\begin{itemize}
    \item \emph{More algorithms.} The most important axis of improvement and future work is to increase the number of algorithms integrated into Flow Gym and their implementation in JAX.
    \item \emph{Volumetric flow-field quantification.} \reviewerThree{Flow Gym currently targets only two-dimensional flow-field quantification, and extending it to volumetric setups is a key next step \cite{annurev:/content/journals/10.1146/annurev-fluid-031822-041721}. Concretely, we plan to introduce an explicit configuration modality that selects the appropriate data interface and estimator pipeline and extend the data container from pairwise image batches (e.g., \texttt{images1}, \texttt{images2}) to \emph{multi-camera grouped batches} for volumetric quantification, where each sample aggregates per-camera image pairs together with camera calibration and pose (intrinsics/extrinsics). Enabling this functionality requires extending the underlying synthetic generator (SynthPix \cite{terpin2025synthpix}) to synthesize \gls*{piv} images in a volumetric setting.}
    \item \emph{Real-world integration.} A promising direction is to connect Flow Gym directly to physical water channels---following efforts such as \cite{terpin2025ff}---potentially via an Internet-accessible interface, enabling algorithms to be evaluated under real-world operating conditions.
\end{itemize}

\bibliographystyle{elsarticle-num} 
\bibliography{references}

\end{document}

%% file: figure_1.tex
\begin{figure}[htb]
    \centering
    \includegraphics[width=\linewidth]{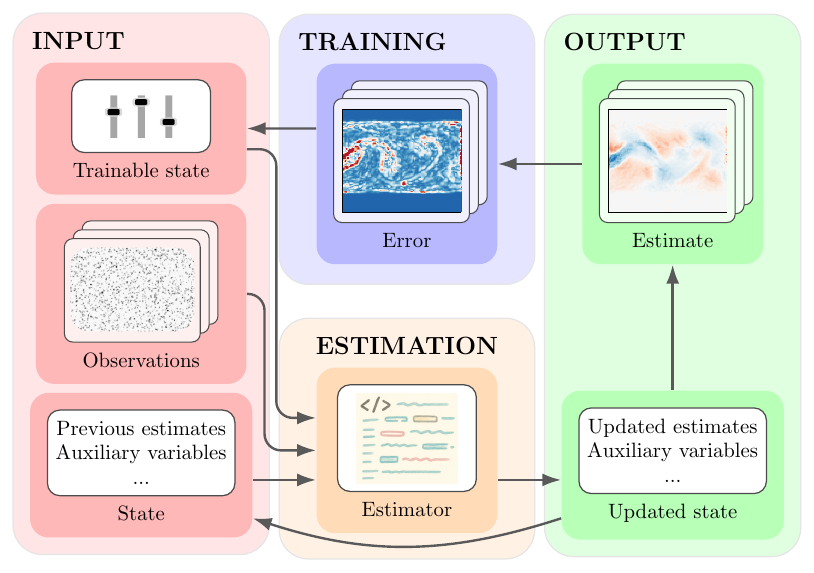}
    \caption{\reviewerOne{Overview of the Flow Gym pipeline centered on the \texttt{Estimator} abstraction. The estimator maps the current observation, the current estimator state, and the trainable state to an estimate and an updated estimator state. The estimator state stores short-term information across calls (e.g., previous estimates and auxiliary variables), while the trainable state contains long-term parameters such as model weights and optimizer state when applicable. During training, the estimate is compared with ground truth or another reference to compute an error used to update the trainable state; during deployment, the same interface applies directly to synthetic or experimental images. Resetting the state enables independent processing of image pairs, whereas carrying it forward enables consecutive estimation on temporal streams.}}
    \label{fig:cover}
\end{figure}

%% file: figure_4.tex
\newcommand{\zoompanel}[2]{%
\begin{minipage}{.32\linewidth}
    \centering
    #2

    \vspace{0.2cm}

    \begin{tikzpicture}
        \node[anchor=south west, inner sep=0] (img) at (0,0)
            {\includegraphics[width=\linewidth]{#1}};

        \begin{scope}[x={(img.south east)},y={(img.north west)}]
            % --- source region ---
            \def\x{0.30}
            \def\y{0.20}
            \def\w{0.08}
            \def\h{0.08}
            
            % --- inset position ---
            \def\ix{0.50}
            \def\iy{0.50}
            \def\iw{0.4}
            \def\ih{0.4}

            % scale factor for the zoomed copy
            \pgfmathsetmacro{\s}{\iw/\w}

            % source rectangle
            \draw[orange, very thick] (\x,\y) rectangle ++(\w,\h);

            % inset frame
            \draw[orange, very thick] (\ix,\iy) rectangle ++(\iw,\ih);

            % optional connector
            \draw[orange, thick] (\x+\w,\y+\h) -- (\ix,\iy);

            % clipped enlarged image inside the inset frame
            \begin{scope}
                \clip (\ix,\iy) rectangle ++(\iw,\ih);
                \node[anchor=south west, inner sep=0]
                    at ({\ix-\s*\x},{\iy-\s*\y})
                    {\includegraphics[width=\s\linewidth]{#1}};
            \end{scope}
        \end{scope}
    \end{tikzpicture}
\end{minipage}%
}
\begin{figure}
    \centering
    \zoompanel{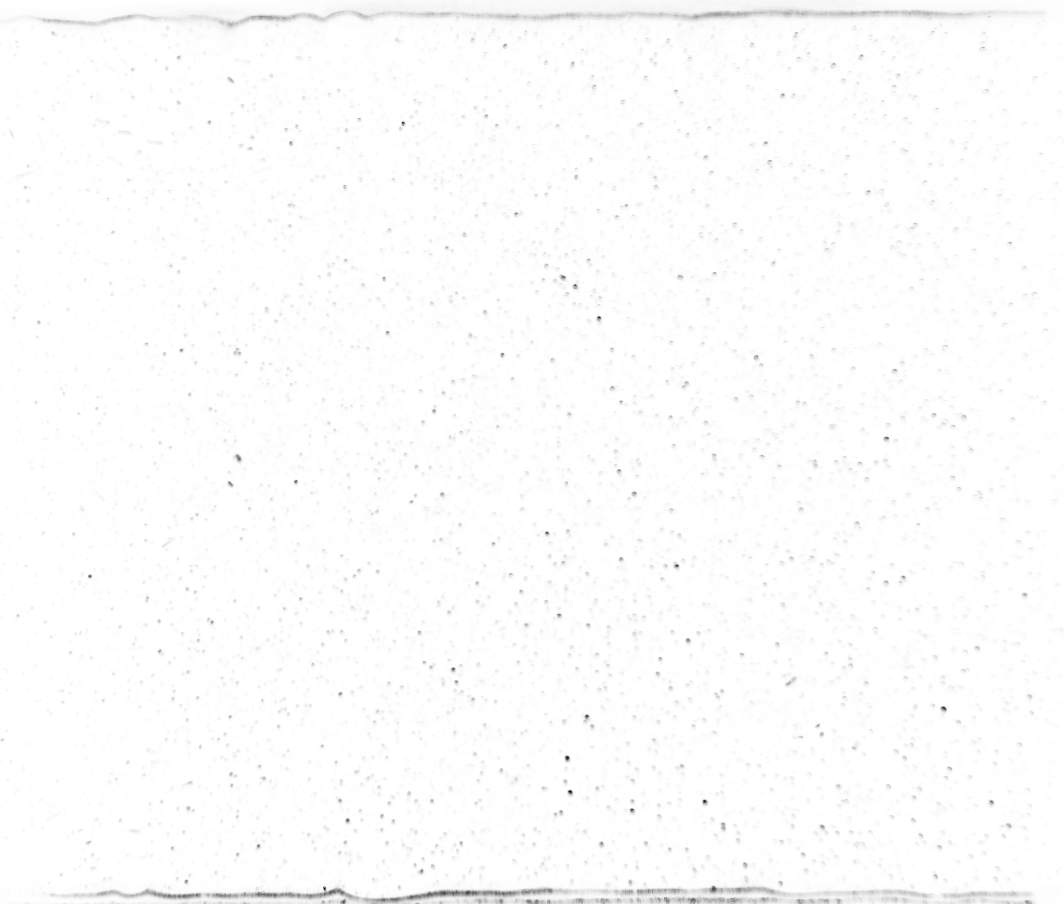}{Original}
    \hfill
    \zoompanel{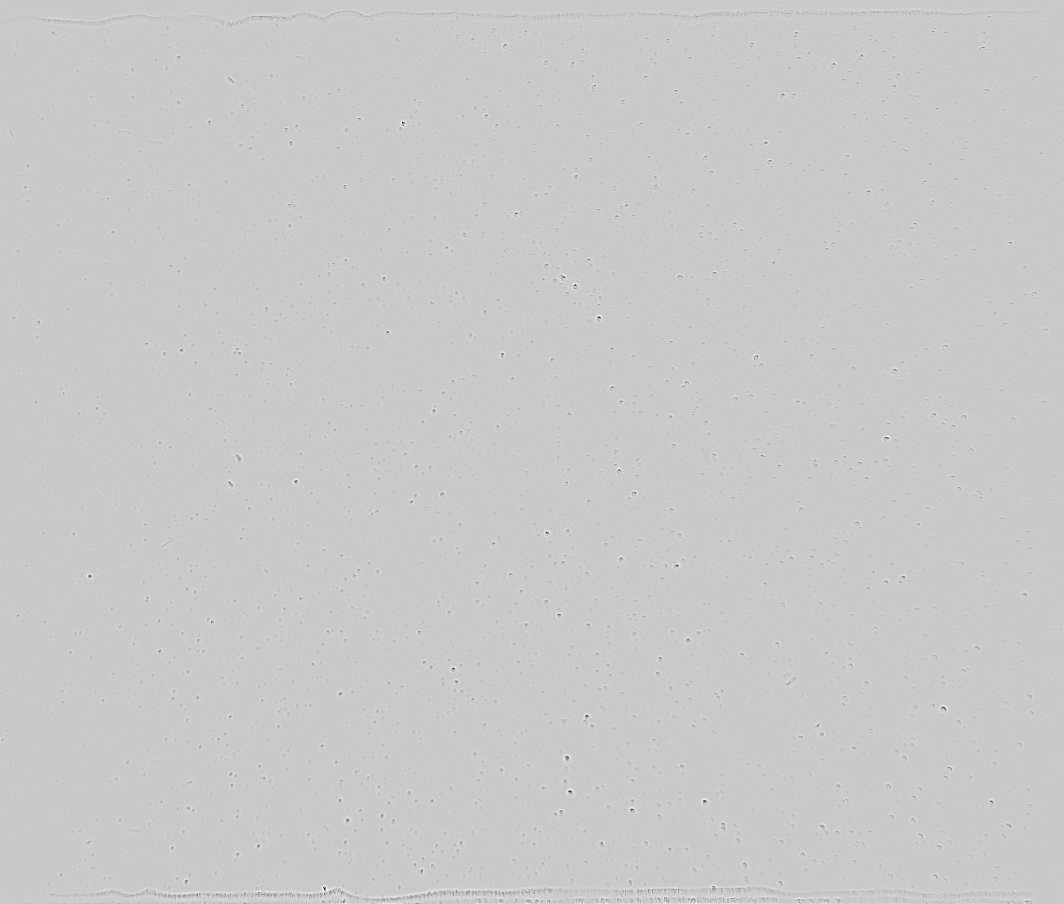}{High-pass}
    \hfill
    \zoompanel{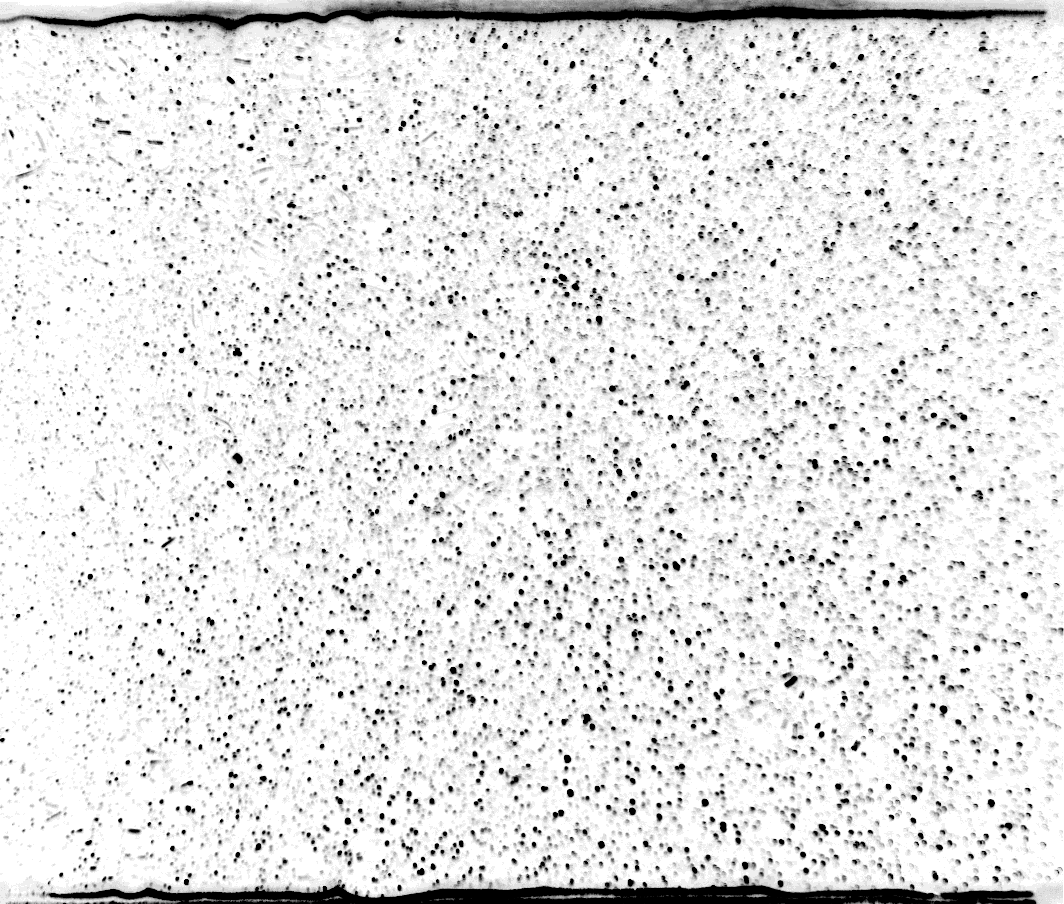}{Intensity clipping}

    \caption{Effects of the pre-processing techniques when applied to images
    from the real setup in \cite{terpin2025ff} ($1064\times904$).
    \reviewerFour{Zoomed insets are provided to highlight the relevant details.}}
    \label{fig:pre-processing}
\end{figure}

%% file: figure_5.tex
\begin{figure}
    \centering
    \begin{minipage}{\linewidth}
        \begin{minipage}{.49\linewidth}
            \centering
            Ground truth
        \end{minipage}
        \hfill
        \begin{minipage}{.49\linewidth}
            \centering
            Raw estimate
        \end{minipage}
    \end{minipage}
    \begin{minipage}{\linewidth}
        \begin{minipage}{.49\linewidth}
            \centering
            \includegraphics[width=\linewidth]{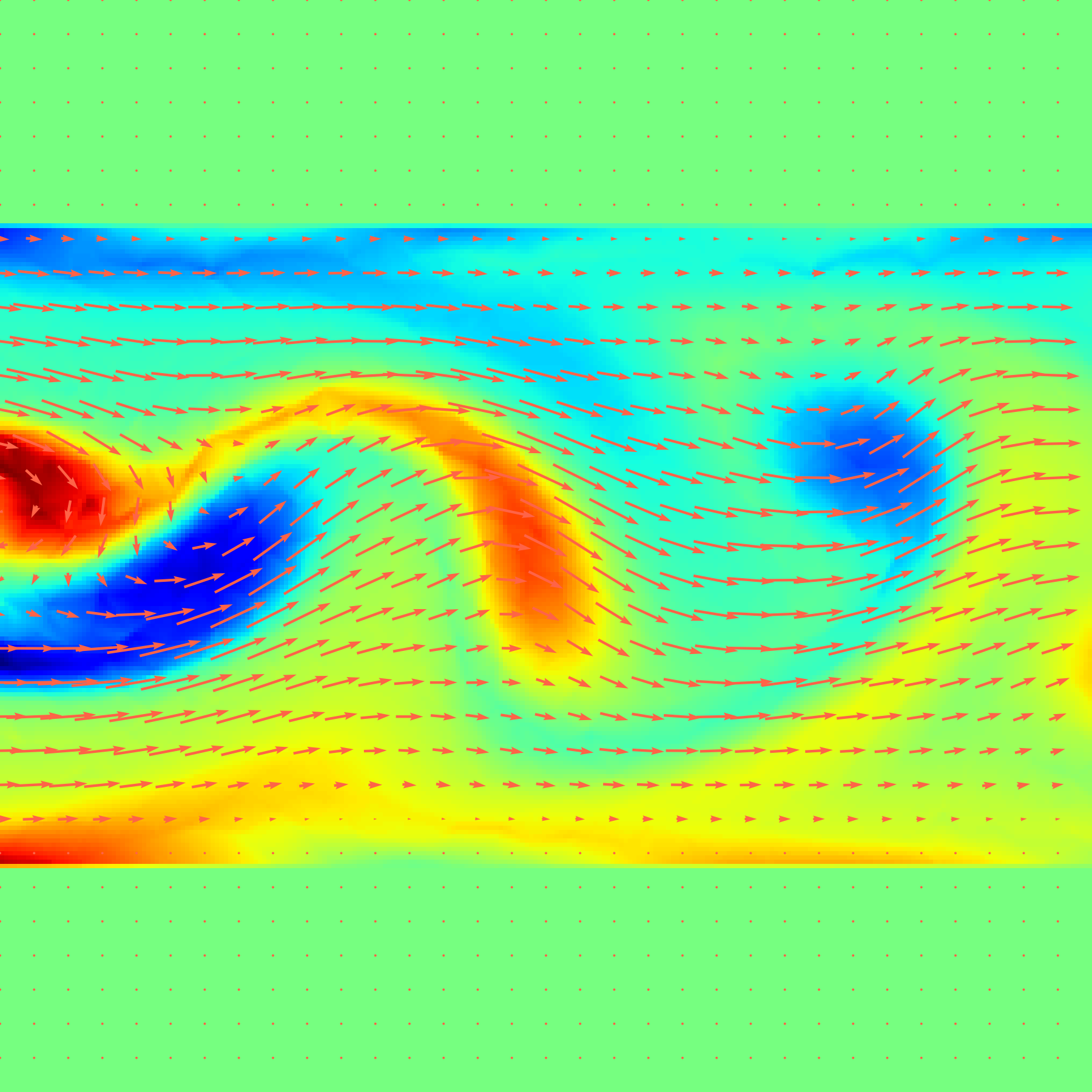}
        \end{minipage}
        \hfill
        \begin{minipage}{.49\linewidth}
            \centering
            \includegraphics[width=\linewidth]{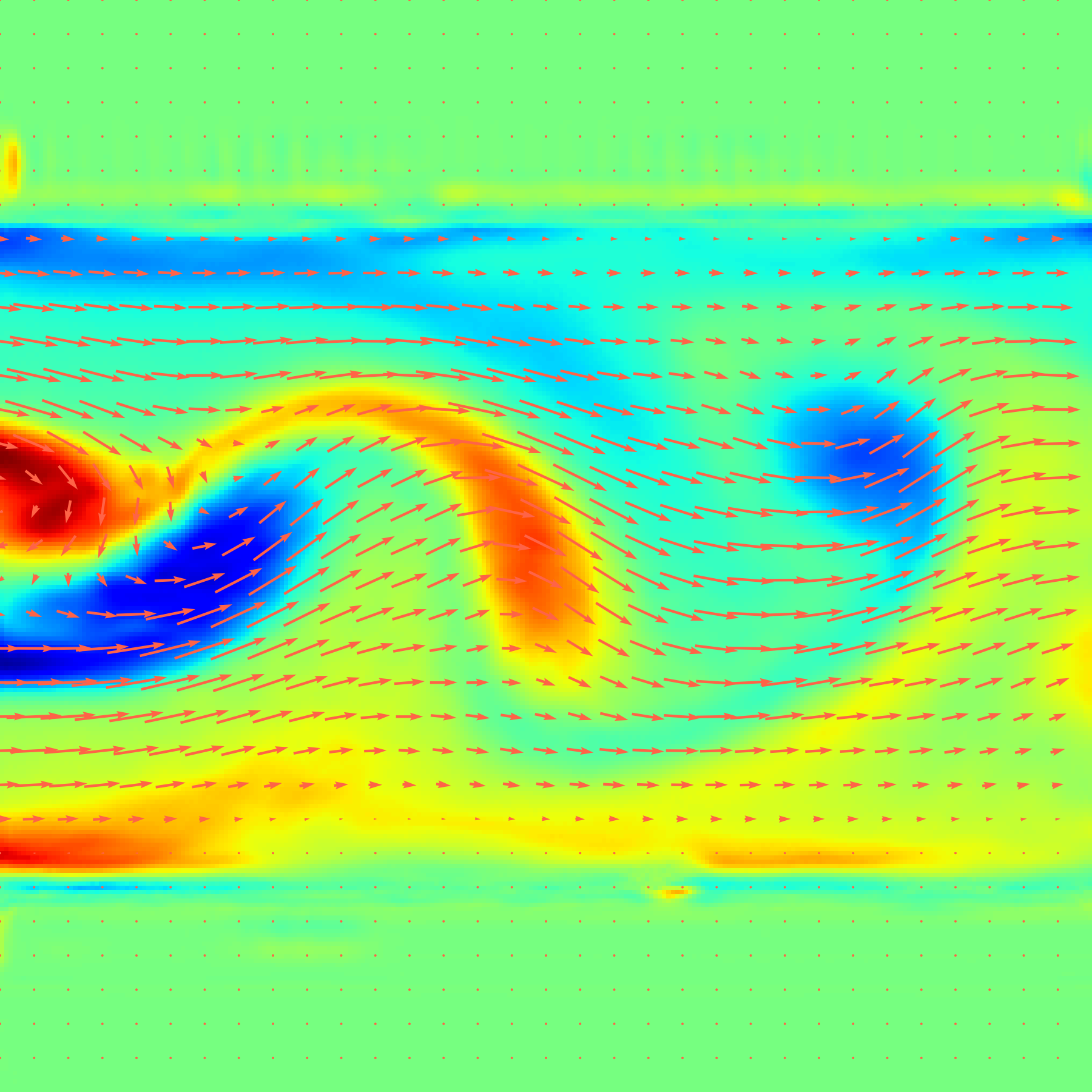}
        \end{minipage}
        \vspace{.5cm}
    \end{minipage}
    \begin{minipage}{\linewidth}
        \begin{minipage}{.32\linewidth}
            \centering
            Local adaptive
        \end{minipage}
        \hfill
        \begin{minipage}{.32\linewidth}
            \centering
            Global adaptive
        \end{minipage}
        \hfill
        \begin{minipage}{.32\linewidth}
            \centering
            Median test
        \end{minipage}
    \end{minipage}
    \begin{minipage}{\linewidth}
        \begin{minipage}{.32\linewidth}
            \centering
           \includegraphics[width=\linewidth]{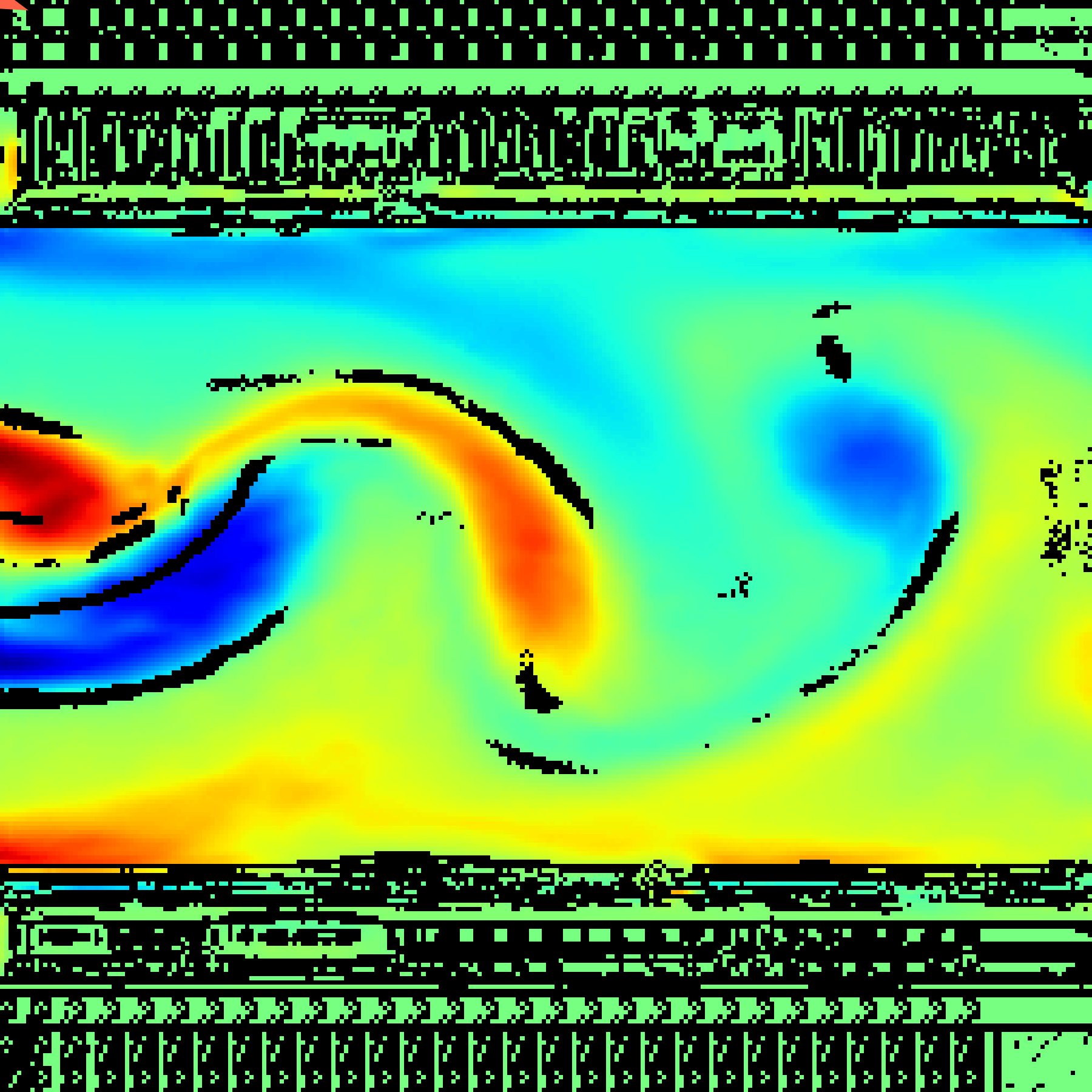}
        \end{minipage}
        \hfill
        \begin{minipage}{.32\linewidth}
            \centering
            \includegraphics[width=\linewidth]{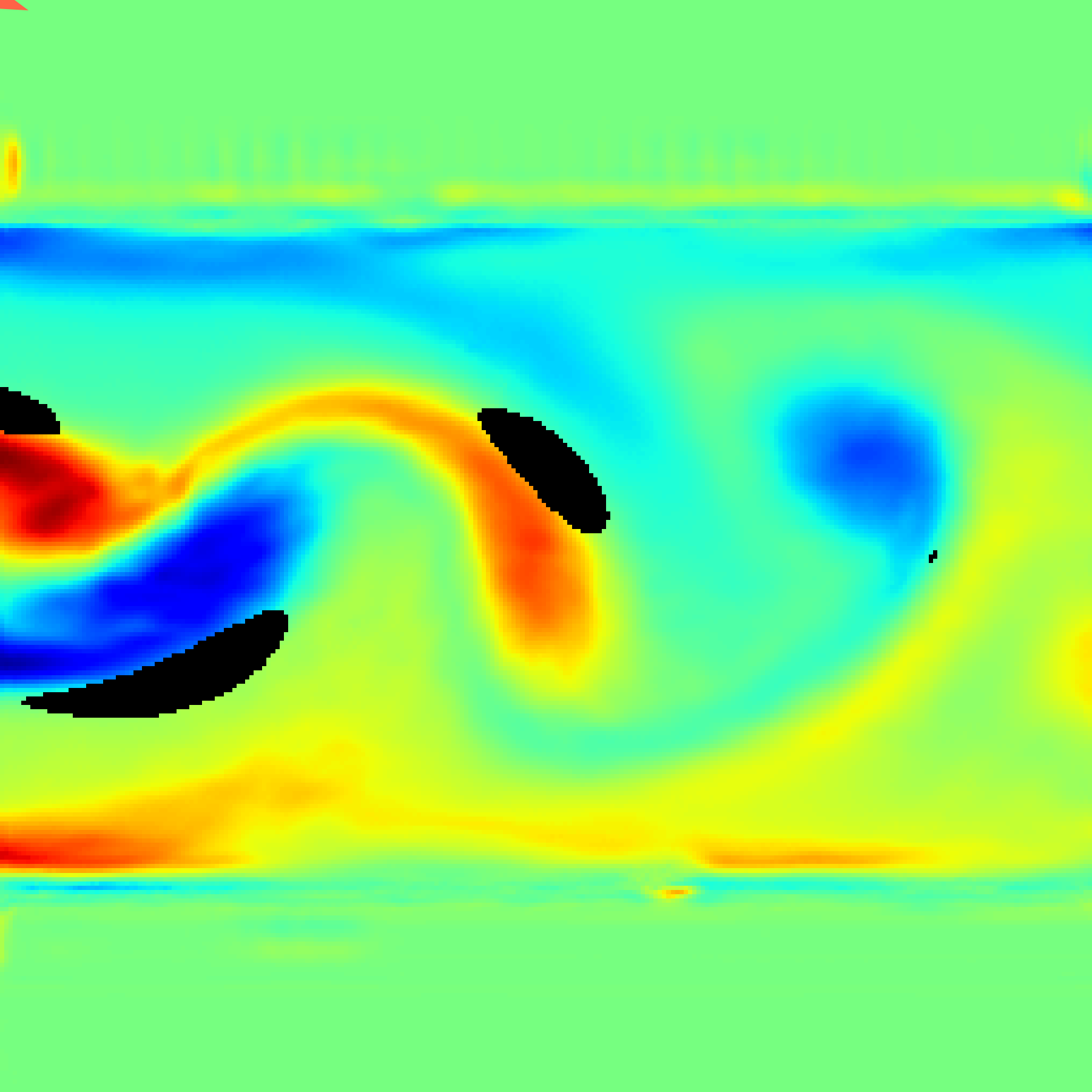}
        \end{minipage}
        \hfill
        \begin{minipage}{.32\linewidth}
            \centering
            \includegraphics[width=\linewidth]{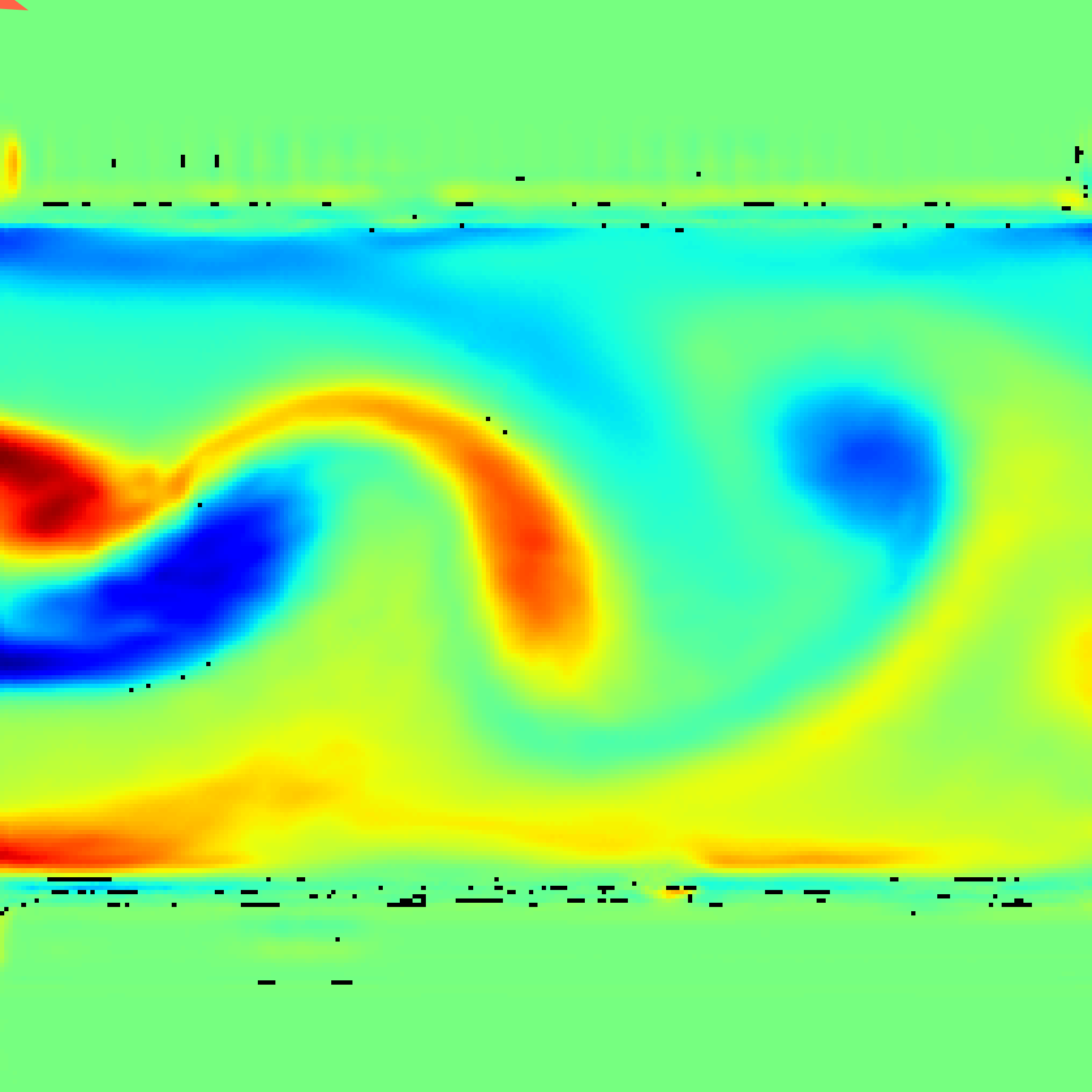}
        \end{minipage}
    \end{minipage}
    \caption{Effects of a selection of the data validation techniques implemented in Flow Gym, when applied to a flow estimated with our implementation in JAX of RAFT32-PIV from a pair of images generated with SynthPix \cite{terpin2025synthpix}. The colormap shows the vorticity of the flow. The RAFT32-PIV neural-network parameters are loaded from the CodeOcean capsule of the original paper \cite{lagemann2021deep}.}
    \label{fig:validation}
\end{figure}